\documentclass[12pt,a4paper]{article}

\usepackage{mathrsfs}
\usepackage{amsmath}
\usepackage{dsfont}
\usepackage{amsfonts}
\usepackage{amssymb}
\usepackage{graphicx}
\usepackage{txfonts}

\newtheorem{defi}{Definition}
\title{Example of a finite game with no Berge equilibria at all}
\author{
  Jaros\l aw Pykacz\\
  Institute of Mathematics, \\ University of Gda\'nsk\\ 
Gda\'nsk, 80-308, Poland\\
\texttt{pykacz@mat.ug.edu.pl}
  \and
  Pawe\l{} Bytner\\
  Institute of Mathematics, \\ University of Gda\'nsk\\ 
Gda\'nsk, 80-308, Poland\\
   \and
  Piotr Fr\c{a}ckiewicz\\
  Institute of Mathematics\\
  Pomeranian University\\
  S\l upsk, 76-200, Poland\\
  \texttt{fracor6@icloud.com}
}
\begin{document}
\maketitle
\begin{abstract}
 The problem of the existence of Berge equilibria in the sense of Zhukovskii in normal form finite games in pure and in mixed strategies is studied. The example of a three-player game that has Berge equilibrium neither in pure nor in mixed strategies is given.
\end{abstract}
\section{Introduction}
The idea of a solution concept that nowadays is usually called {\it Berge equilibrium} ({\em in the sense of Zhukovskii}) was launched by the French mathematician Claude Berge in his book {\it Th\'{e}orie g\'{e}n\'{e}rale des jeux \`{a} n personnes} \cite{Ber57}. Unfortunately this book remained virtually unnoticed in the English-speaking world. However, Berge's book was translated into Russian in 1961 and, maybe because the very idea of Berge equilibrium fits very well to the most basic idea of communist ideology, it gained remarkable popularity in the former Soviet Union. In particular, numerous students from Arabic countries that studied there got acquainted with this idea and, after coming back from the Soviet Union, they continued studies on this valuable idea and returned it to the French- and English-speaking world\footnote{Brief description of historical development of the idea of Berge equilibrium can be found in a paper \cite{CKMT11}. See also very extensive review of literature on Berge equilibria \cite{LZ17}.}

Actually, the idea of Berge equilibrium is in a sense opposite to the idea of Nash equilibrium. While Nash equilibrium is based on egoism: each player aims to maximize his own payoff, Berge equilibrium is based on altruism: each player's aim is to maximize payoffs of all the other players, so when every player does so, everyone is better off. This explains why, in general, Berge equilibria yield to players higher payoffs than Nash equilibria.

However, it seems that Nash equilibria outperform Berge equilibria  in one respect: the historic theorem stated in \cite{Nas50} assures that in any finite game there exists at least one Nash equilibrium - in pure or mixed strategies. In the case of Berge equilibria this problem is still a subject of a debate. The aim of this paper is to give an example of a three-player game that has  no Berge equilibria at all: neither in pure nor in mixed strategies. The importance of this example follows from two facts: (i) According to the results of \cite{CKMT11} extended to mixed strategies, in a finite two-player game  Nash equilibria in the original game  and Berge equilibria in a game with interchanged payoffs are in 1--1 correspondence. Therefore, due to the Nash theorem \cite{Nas50}, every two-player finite game has at least one Berge equilibrium -- in pure or mixed strategies. The problem whether we should expect the same in games with a bigger number of players was posed as the Open Problem No. 1 in \cite{LZ17}. Our counterexample shows that already among three-person games there exist games with no Berge equilibria at all. (ii) Zhukovskii et al., \cite{ZSS14} proved the existence of Berge equilibrium (in pure or mixed strategies) in $n$-player games with compact sets of strategies and continuous payoff functions, however, with additional constraints. Our counterexample shows that this result cannot be extended to the class of the most interesting games: $n$-player finite games.

The studies of proposed solutions to the problem of the existence of Berge equilibria in pure and mixed strategies are a little bit difficult because it is rather rare that the authors declare openly, like in the papers \cite{CKMT11} and \cite{CK15}, that they study these equilibria in pure strategies only. More common is the attitude encountered, e.g., in the paper \cite{MPT12} that only after careful reading of these papers a reader can find some fragments of a text from which he can infer that these papers deal with equilibria in pure strategies only\footnote{Take a look on the line 10 from the top on the page 1250005-5 of the paper \cite{MPT12}. The authors write: ``{\it The unique Nash equilibrium is (D,E)}" while actually in this game there are also numerous Nash equilibria in mixed strategies. From this a careful reader can infer that the authors deal exclusively with equilibria in pure strategies, although they do not write this explicitly.}. This is a serious drawback of numerous papers on Berge equilibria and the present authors appeal to all the scholars active in this field  to make this issue clear in the very beginning of all the papers they write.
\section{Basic Notions and Definitions}
Noncooperative finite game in normal  form is a triple:
\begin{equation}
G = \langle N, (S_i)_{i\in N}, (U_i)_{i\in N}\rangle ,
\end{equation}
where $N = \{1,\ldots ,n\}$ denotes the set of players, $S_i$ is a finite set of pure strategies of a player $i$, and $U_i$ is a function from $S = \prod_{i\in N}S_i $ into the set of real numbers that describes payoffs possible to obtain by the player $i$. Mixed strategy of the player $i$ is identified with a probability distribution defined on the set $S_i$ of his pure strategies. The set of all mixed strategies of the player $i$ is denoted $\widetilde{S}_i$. When at least one player chooses a ``genuine mixed" (i.e., non-pure) strategy, payoffs are understood as suitable expected values, and the real-valued function they form, defined on $\widetilde{S} = \prod_{i\in N}\widetilde{S}_i $,  will be denoted $\widetilde{U}_i$.  We do not distinguish between a game and its mixed extension, and when we write {\it strategy} we mean a general mixed strategy, with pure strategies being special cases of mixed ones. Let ${\bf s} = (s_1,\ldots ,s_n) \in  \prod_{i\in N}\widetilde{S}_i $ be a {\it strategy profile}, then by ${\bf s}_{-i}$ we denote the {\it incomplete strategy profile} ${\bf s}_{-i} = (s_1,\ldots s_{i-1}, s_{i+1},\ldots ,s_n)\in \widetilde{S}_{-i} =  \prod_{j \neq i}\widetilde{S}_j $. By a small abuse of symbols we make an identification $(s_i , {\bf s}_{-i}) = {\bf s}$.

\begin{defi}
{\em A strategy profile ${\bf s}^* = (s^*_1, \ldots ,s^*_n)  \in \widetilde{S}$ is a {\em Berge equilibrium} (in the sense of Zhukovskii) of the game $G$ if:}
\begin{equation}
\forall i \in N, \quad  \forall {\bf s}_{-i} \in \widetilde{S}_{-i}, \quad \widetilde{U}_i(s^*_i,{ \bf s}_{-i}) \leq \widetilde{U}_i({\bf s}^*).
\end{equation}
\end{defi}
Let us compare this notion with the notion of Nash equilibrium:

\begin{defi}
{\em A strategy profile ${\bf s}^* = (s^*_1, \ldots ,s^*_n) \in \widetilde{S}$ is a {\em Nash equilibrium} of the game $G$ if:}
\begin{equation}
\forall i \in N, \quad  \forall s_{i} \in \widetilde{S}_{i}, \quad \widetilde{U}_i(s_i,{ \bf s}^*_{-i}) \leq \widetilde{U}_i({\bf s}^*).
\end{equation}
\end{defi}
In small games all Nash equilibria can be easily found as intersections of graphs of the best reply correspondences (see, for example, \cite{Pet08}). Musy et al. in the work \cite{MPT12} introduced an analogous notion of the best support correspondence which enables one to find, at least in small games, all Berge equilibria. Although, as we mentioned, after a careful reading of their  paper it becomes clear that they deal with Berge equilibria in pure strategies only, their results can be in a straightforward way extended to mixed strategies. Therefore, we adopt the following definition that extends to mixed strategies the original definition \cite{MPT12}.

\begin{defi}
{\em Let $i \in N$, $s_i \in \widetilde{S}_i$. An incomplete strategy profile $\overline{\bf{s}}_{-i} \in \widetilde{S}_{-i}$ is  {\em the best support} to strategy $s_i$ if:
\begin{equation}
\forall {\bf s}_{-i} \in \widetilde{S}_{-i} \qquad \widetilde{U}_i (s_i,  {\bf s}_{-i}) \leq \widetilde{U}_i (s_i,\overline{\bf{s}}_{-i}).
\end{equation}
}
\end{defi}

It was noted in \cite{MPT12} that the best support correspondences can be used to reformulate the definition of Berge equilibrium in the same way as the best reply correspondences enable us to reformulate the definition of Nash equilibrium:

``{\em A strategy profile is a Nash equilibrium if and only if each player's strategy is a best reply to the co-players' incomplete strategy profile, whereas a strategy profile is a Berge equilibrium if and only if co-payers' incomplete strategy profile is a best support to each player's strategy}".

It is straightforward to see that this refers also to Nash and Berge equilibria in mixed strategies.

\section{Example of a 3-Player Game with no Berge Equilibria at all}

Let us consider the following 3-player game in which each of the players has two pure strategies. Pure strategies of the first, the second, and the third player are denoted $A_1, A_2$; $B_1, B_2$; $C_1, C_2$, respectively.
\begin{equation}\label{game}
 C_{1}\colon~ \bordermatrix{& B_1 & B_2 \cr 
A_1 & (2,1,0) & (1,1,1) \cr 
A_2  & (2,0,1) & (1,0,2)} \quad  ~C_{2}\colon~ \bordermatrix{& B_1 & B_2 \cr 
A_1 & (1,2,0) & (0,2,1) \cr 
A_2  & (1,1,1) & (0,1,2)}.
\end{equation}
The left-hand matrix refers to the pure strategy $C_1$ of the third player, while the right-hand matrix refers to his pure strategy $C_2$. Let us note that this game is a very special one: None of the players has any possibility to influence his own payoff, no matter if he uses any of his pure or mixed strategies. On the contrary, his payoffs depend exclusively on the choices of the remaining players. 
\begin{figure}
\centering
  \includegraphics[width=1\textwidth]{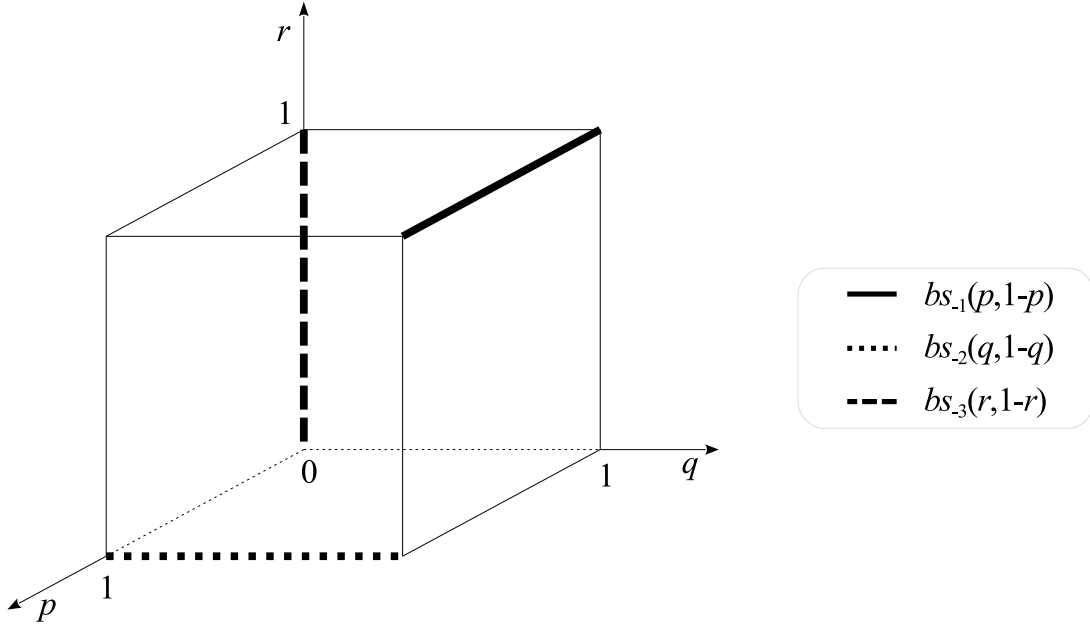}
\caption{Best support correspondences $bs_{i}(\cdot)$ corresponding to game (\ref{game}).\label{fig1}}
\label{fig1}       
\end{figure}
This very special feature implies that any strategy profile of this game, consisting of strategies of any type: pure, genuine mixed, or both, is a (weak) Nash equilibrium. Moreover, since this game is a constant-sum game, increasing the payoff of one player inevitably leads to decreasing the payoffs of some, or all, other players, so any triple of payoffs yielded by any strategy profile is optimal in Pareto sense. Therefore, neither the concept of Nash equilibrium, nor the concept of paretooptimal results provides any hint to players which strategies should be chosen. Sometimes it is claimed (cf. \cite{LZ17}) that in such cases the concept of Berge equilibrium could be of some help. So, let us study Berge equilibria in this game.

One can easily check that the second and the third players' best support to any of the first player (pure or mixed) strategy is a pair of pure strategies $(B_1,C_1)$, the first and the third players' best support to any of the second player (pure or mixed) strategy is a pair of pure strategies $(A_1, C_2)$, and finally the first and the second players' best support to any of (pure or mixed) strategy of the third player is a pair of pure strategies $(A_2,B_2)$. 
More formally, let us denote by $p,q$ and $r$ the probabilities with which the first, the second, and the third player choose, respectively, their first pure strategy. Denote by $bs_{-i}(s_{i})$ the set of best supports $\overline{\bf{s}}_{-i}$ to strategy $s_{i}$.
This means that if we plot graphs of the best support correspondences $bs_{-i}(\cdot)$ as subsets of the 3-dimensional unit cube $[0,1]^3$, these graphs form, respectively, edges $(p\in [0,1], 1,1)$, $(1,q\in [0,1], 0)$, and $(0,0,r\in [0,1])$ of this cube. Their intersection is an empty set, so this game has no Berge equilibria either in pure or in mixed strategies (see Fig. 1).

Let us note that the game (5) was considered for the first time in a MSc. Thesis of P. Bytner \cite{Byt16}, where he proved that this game has no Berge equilibria, either in pure or in mixed strategies, using another methods. These methods will be described in a subsequent paper.

\section*{Acknowledgments}

Work by Piotr Fr\c{a}ckiewicz and Jaros\l aw Pykacz was supported by the National Science Centre, Poland under the research project 2016/23/D/ST1/01557.


\begin{thebibliography}{100}
\bibitem{Ber57} Berge, C. {\em Th\'{e}orie g\'{e}n\'{e}rale des jeux \`{a} n personnes}; Gauthier-Villars, Paris, 1957.
\bibitem{CKMT11} Colman, A. M. Mutual support in games: Some properties of Berge equilibria. {\em J. Math. Psychol.} {\bf 2011}, {\em 55}, 166.
\bibitem{Byt16} Bytner, P. Berge Equilibria in Pure and Mixed Strategies. MSc. Thesis (in Polish) (University of Gda\'nsk, Gda\'nsk) 2016.
\bibitem{CK15} Corley, H. W.;  Kwain, P. An algorithm for computing all Berge equilibria. {\em Game Theory} {\bf 2015}, Article ID 862842.
\bibitem{LZ17} Larbani, M.; Zhukovskii, V. I. Berge equilibrium in normal form static games: A literature review. {\em Izv, Inst. Mat. Inform. Udmurt. Gos. Univ.} {\bf 2017}, {\em 49}, 80.
\bibitem{MPT12} Musy, O.; Pottier, A.; Tazda{\"{i}}t, T. A new theorem to find Berge equilibria. {\em Int. Game Theory Rev.} {\bf 2012}, {\em 14}, 1250005.
\bibitem{Nas50} Nash, J. Equilibrium points in $n$-person games {\em Proc. Natl. Acad. Sci. U.S.A.} {\bf 1950}, {\em 36}, 48.
\bibitem{Pet08} Peters, H. Game Theory. A Multi-Leveled Approach, Springer, Berlin, 2008.
\bibitem{ZSS14} Zhukovskii, V. I.; Sachkov, S. N.; Smirnova L. V. Existence of Berge equilibrium in mixed strategies {\em Uchenye Zapiski Tavricheskogo Natsional'nogo Universiteta im. V. I. Vernadskogo} {\bf 2014}, {\em 27}, 261. 
\end{thebibliography}
\end{document}